\documentclass[aps,prl,twocolumn,showpacs,groupedaddress]{revtex4}
\usepackage{amssymb}
\usepackage{amsmath}
\usepackage{graphicx}

\begin{document}

\title{Separability Criterion of Tripartite Qubit Systems }
\author{Chang-shui Yu}
\email{quaninformation@sina.com}
\author{He-shan Song}
\affiliation{Department of Physics, Dalian University of Technology,\\
Dalian 116024, China}
\date{\today }

\begin{abstract}
In this paper, we present a method to construct full separability criteria
for tripartite systems of qubits. The spirit of our approach is that a
tripartite pure state can be regarded as a three-order tensor that provides
an intuitionistic mathematical formulation for the full separability of pure
states. We extend the definition to mixed states and given out the
corresponding full separability criterion. As applications, we discuss the
separability of several bound entangled states, which shows that our
criterion is feasible.
\end{abstract}

\maketitle

\section{\protect\bigskip Introduction}

Entanglement is an essential ingredient in quantum information and the
central feature of quantum mechanics which distinguishes a quantum system
from its classical counterpart. In recent years, it has been regarded as an
important physical resource, and widely applied to a lot of quantum
information processing(QIP): quantum computation [1], quantum cryptography
[2], quantum teleportation [3], quantum dense coding [4] and so on.

Entanglement arises only if some subsystems ever interacted with the others
among the whole multipartite system in physics, or only if the multipartite
quantum state is not separable or factorable in mathematics. The latter
provides a direct way to tell whether or not a given quantum state is
entangled.

As to the separability of bipartite quantum states, partial entropy
introduced by Bennett \textit{et} \textit{al}. [5] provides a good criterion
of separability for pure states. Later, Wootters presents the remarkable
concurrence for bipartite systems of qubits [6,7]. Based on the motivation
of generalizing the definition of concurrence to higher dimensional systems,
many attempts have been made [8,9,10,11], which provide good separability
criteria\ for bipartite qubit systems under corresponding conditions, whilst
Ref.[8] also presents an alternative method to minimize the convex hull for
mixed states. As to multipartite quantum systems, several separability
criteria have been proposed [12,13,14,15,16,17]. The most notable one is
3-tangle for three qubits [13]. Recently, the result has been generalized to
the higher dimensional systems [18]. Despite the enormous effort, the
separability of quantum states especially in higher dimensional systems is
still an open problem.

In this paper we construct the full separability criteria for arbitrary
tripartite qubit system by a novel method, i.e. a tripartite pure state can
be defined by a three-order tensor. The definition provides an
intuitionistic mathematical formulation for the full separability of pure
states. Analogous to Ref.[8], we extend the definition to mixed states. More
importantly, Our approach is easily generalized to higher dimensional
systems. As applications, we discuss separability of two bound entangled
states introduced in [19,20], respectively.

\section{\protect\bigskip Separability for pure states}

We start with the separability definition of tripartite qubit pure state $%
\left\vert \psi \right\rangle _{ABC}.$ $\left\vert \psi _{ABC}\right\rangle $
is fully separable iff
\begin{equation}
\left\vert \psi \right\rangle _{ABC}=\left\vert \psi \right\rangle
_{A}\otimes \left\vert \psi \right\rangle _{B}\otimes \left\vert \psi
\right\rangle _{C}.
\end{equation}%
Consider a general tripartite pure state written by%
\begin{equation}
\left\vert \psi \right\rangle _{ABC}=\sum a_{ijk}\left\vert i\right\rangle
_{A}\left\vert j\right\rangle _{B}\left\vert k\right\rangle _{C},
\end{equation}%
where $i$, $j$, $k=0$, $1$, the coefficients $a_{ijk}$s can be arranged as a
three-order tensor (tensor cube) [21] as shown in figure 1.
\begin{figure}[tbp]
\includegraphics[width=8.5cm]{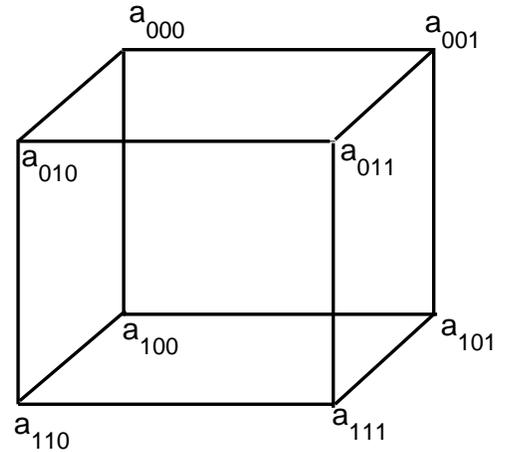}
\caption{Three-order tensor of the coefficients of a tripartite pure state.}
\label{1}
\end{figure}

Note that the subscripts of $a_{ijk}$ correspond to the basis $\left\vert
i\right\rangle _{A}\left\vert j\right\rangle _{B}\left\vert k\right\rangle
_{C}$. Every surface can be regarded as the tensor product of a single qubit
and an unnormalized bipartite state. Hence, if the two vectors (edges) of a
surface are linear relevant (including one of the vectors is $zero$ vector),
then the bipartite state mentioned above can be factorized. The conclusion
for diagonal plane is analogous. Considering all the planes, one can easily
find that the tripartite state is fully separable iff all the vectors which
are parallel mutually shown in the cube, are linear relevant, according to
the fundamental linear algebra. I.e. the rank of every matrix composed of
four coefficients on the corresponding surface and diagonal plane is $one$.
Equivalently, we can obtain the following lemma.

\textbf{Lemma1.-}A tripartite pure state $\left\vert \psi \right\rangle
_{ABC}$ with the form of eq.(2) in $2\times 2\times 2$ dimensional Hilbert
space is fully separable, iff the following six equations hold:
\begin{equation}
\sum\limits_{i=0}^{1}\left\vert \left( a_{i00}a_{i11}-a_{i01}a_{i10}\right)
\right\vert =0,
\end{equation}%
\begin{equation}
\sum\limits_{i=0}^{1}\left\vert \left( a_{0i0}a_{1i1}-a_{0i1}a_{1i0}\right)
\right\vert =0,
\end{equation}%
\begin{equation}
\sum\limits_{i=0}^{1}\left\vert \left( a_{00i}a_{11i}-a_{01i}a_{10i}\right)
\right\vert =0,
\end{equation}%
\begin{equation}
\sum\limits_{i=0}^{1}\left( a_{0i0}a_{1j1}-a_{0j1}a_{1i0}\right) =0,
\end{equation}%
\begin{equation}
\sum\limits_{i=0}^{1}\left( a_{i00}a_{j11}-a_{j01}a_{i10}\right) =0,
\end{equation}%
\begin{equation}
\sum\limits_{i=0}^{1}\left( a_{00i}a_{11j}-a_{01j}a_{10i}\right) =0.
\end{equation}%
where $i,j=0,1$ and $i\oplus j=1$.

\bigskip \textbf{Proof.} (Sufficient condition) If eq.(3)-eq.(5) hold, then
the rank of every matrix that the cubic surface corresponds to is \textit{one%
}. If eq.(6)-eq.(8) hold, then the rank of every matrix that the cubic
diagonal plane corresponds to is \textit{one}. Hence, That eq.(3)-eq.(8)
hold simultaneously shows that the tripartite pure state can be fully
factorized, i.e. it is fully separable.

(Necessary condition) If a given tripartite state is fully separable, one
can easily obtain that the rank of every corresponding matrix is $one$.
Namely, eq.(3)-eq.(8) hold.

Consider that $\left\vert \psi \right\rangle _{ABC}$ can be denoted by a
vector in $2\times 2\times 2$ dimensional Hilbert space,%
\begin{equation*}
\left\vert \psi \right\rangle
=(a_{000},a_{001},a_{010},a_{011},a_{100},a_{101},a_{110,}a_{111})^{T},
\end{equation*}%
with the superscript $T$ denoting transpose, we can write the above
equations (3-8) in matrix notation by%
\begin{equation*}
\left\langle \psi ^{\ast }\right\vert s^{\alpha }\left\vert \psi
\right\rangle =0,\text{ }\alpha =1,2,\cdot \cdot \cdot ,9,
\end{equation*}%
where the star denotes complex conjugation, and $s^{1}=-\sigma _{y}\otimes
\sigma _{y}\otimes I_{1}$, $s^{2}=-\sigma _{y}\otimes \sigma _{y}\otimes
I_{2}$, $s^{3}=-\sigma _{y}\otimes I_{1}\otimes \sigma _{y}$, $s^{4}=-\sigma
_{y}\otimes I_{2}\otimes \sigma _{y}$, $s^{5}=-I_{1}\otimes \sigma
_{y}\otimes \sigma _{y}$, $s^{6}=-I_{2}\otimes \sigma _{y}\otimes \sigma
_{y} $, $s^{7}=-Iv\otimes \sigma _{y}\otimes \sigma _{y}$, $s^{8}=-\sigma
_{y}\otimes Iv\otimes \sigma _{y}$, $s^{9}=-\sigma _{y}\otimes \sigma
_{y}\otimes Iv$, with $\sigma _{y}=\left(
\begin{array}{cc}
0 & -i \\
i & 0%
\end{array}%
\right) $, $I_{1}=\left(
\begin{array}{cc}
1 & 0 \\
0 & 0%
\end{array}%
\right) $, $I_{2}=\left(
\begin{array}{cc}
0 & 0 \\
0 & 1%
\end{array}%
\right) $ and $Iv=\left(
\begin{array}{cc}
0 & 1 \\
1 & 0%
\end{array}%
\right) $. (Alternatively, $s^{1}\ldots s^{6}$ can be replaced by $%
S^{1}\cdots S^{3}$ for tripartite pure states of qubit with $S^{1}=-\sigma
_{y}\otimes \sigma _{y}\otimes \tilde{I}$, $S^{2}=-\sigma _{y}\otimes \tilde{%
I}\otimes \sigma _{y}$ and $S^{3}=-\tilde{I}\otimes \sigma _{y}\otimes
\sigma _{y}$, where $\tilde{I}=\left(
\begin{array}{cc}
1 & 0 \\
0 & i%
\end{array}%
\right) $, by which $3$ comlex optimal parameters can be reduced for the
case of mixed states).

Define a new vector $\boldsymbol{C}(\psi )$ by $\boldsymbol{C}(\psi )=%
\overset{9}{\underset{\alpha =1}{\oplus }}C^{\alpha }$ with $C^{\alpha }=$ $%
\left\langle \psi ^{\ast }\right\vert s^{\alpha }\left\vert \psi
\right\rangle $, then the length of the vector can be given by $\left\vert
\boldsymbol{C}(\psi )\right\vert =\sqrt{\sum\limits_{\alpha }\left(
C^{\alpha }\right) ^{2}}$. Therefore, the full separability criterion for a
tripartite state can be expressed by a more rigorous form as follows.

\textbf{Theorem1.-}A tripartite pure state $\psi $ is fully separable iff $%
\left\vert \boldsymbol{C}(\psi )\right\vert =0$.

\textbf{Proof. }That\textbf{\ }$\left\vert \boldsymbol{C}(\psi )\right\vert
=0$ is equivalent to that $C^{\alpha }=0$ holds for any $\alpha $. According
to Lemma 1, one can obtain that $\left\vert \boldsymbol{C}(\psi )\right\vert
=0$ is the sufficient and necessary condition.

\section{Separability for mixed states}

A tripartite mixed state $\rho $ is fully separable iff there exists a
decomposition $\rho =\sum\limits_{k=1}^{K}\omega _{k}\left\vert \psi
^{k}\right\rangle \left\langle \psi ^{k}\right\vert $, $\omega _{k}>0$ such
that $\psi ^{k}$ is fully separable for every $k$ or equivalently iff the
infimum of the average $\left\vert \boldsymbol{C}(\psi ^{k})\right\vert $
vanishes, namely,%
\begin{equation}
C(\rho )=\inf \sum\limits_{k}^{K}\omega _{k}\left\vert \boldsymbol{C}(\psi
^{k})\right\vert =0,
\end{equation}%
among all possible decompositions. Therefore, for any a given decomposition
\begin{equation}
\rho =\sum\limits_{k=1}^{K}\omega _{k}\left\vert \psi ^{k}\right\rangle
\left\langle \psi ^{k}\right\vert ,
\end{equation}%
according to the Minkowski inequality%
\begin{equation}
\left( \sum\limits_{i=1}\left( \sum\limits_{k}x_{i}^{k}\right) ^{p}\right)
^{1/p}\leq \sum_{k}\left( \sum\limits_{i=1}\left( x_{i}^{k}\right)
^{p}\right) ^{1/p},\text{ }p>1,
\end{equation}%
one can get%
\begin{eqnarray}
C(\rho ) &=&\inf \sum\limits_{k}\omega _{k}\left\vert \boldsymbol{C}(\psi
^{k})\right\vert  \notag \\
&=&\inf \sum\limits_{k}\omega _{k}\sqrt{\sum\limits_{\alpha }\left\vert
\left\langle (\psi ^{k})^{\ast }\right\vert s^{\alpha }\left\vert \psi
^{k}\right\rangle \right\vert ^{2}}  \notag \\
&\geq &\inf \sqrt{\sum\limits_{\alpha }\left( \sum\limits_{k}\omega
_{k}\left\vert \left\langle (\psi ^{k})^{\ast }\right\vert s^{\alpha
}\left\vert \psi ^{k}\right\rangle \right\vert \right) ^{2}}.
\end{eqnarray}%
Consider the matrix notation [8] of equation (10) as $\rho =\Psi W\Psi
^{\dagger }$, where $W$ is a diagonal matrix with $W_{kk}=\omega _{k}$, the
columns of \ the matrix $\Psi $ correspond to the vectors $\psi ^{k}$, and
the eigenvalue decomposition, $\rho =\Phi M\Phi ^{\dagger }$, where $M$ is a
diagonal matrix whose diagonal elements are the eigenvalues of $\rho $, and $%
\Phi $ is a unitary matrix whose columns are the eigenvectors of $\rho $,
associated with the relation $\Psi W^{1/2}=\Phi M^{1/2}U$, where $U$ is a
Right-unitary matrix, inequality (12) can be rewritten as
\begin{eqnarray}
C(\rho ) &\geq &\inf \sqrt{\sum\limits_{\alpha }\left(
\sum\limits_{k}\left\vert \Psi ^{T}W^{1/2}s^{\alpha }W^{1/2}\Psi \right\vert
_{kk}\right) ^{2}}  \notag \\
&=&\underset{U}{\inf }\sqrt{\sum\limits_{\alpha }\left(
\sum\limits_{k}\left\vert U^{T}M^{1/2}\Phi ^{T}s^{\alpha }\Phi
M^{1/2}U\right\vert _{kk}\right) ^{2}}.
\end{eqnarray}%
In terms of the Cauchy-Schwarz inequality
\begin{equation}
\left( \sum\limits_{i}x_{i}^{2}\right) ^{1/2}\left(
\sum\limits_{i}y_{i}^{2}\right) ^{1/2}\geqslant \sum\limits_{i}x_{i}y_{i},
\end{equation}%
the inequality
\begin{equation}
C(\rho )\geq \underset{U}{\inf }\sum\limits_{k}\left\vert U^{T}\left(
\sum\limits_{\alpha }z_{\alpha }A^{\alpha }\right) U\right\vert _{kk}
\end{equation}%
is implied for any $z_{\alpha }=y_{\alpha }e^{i\phi }$ with $y_{\alpha }>0$
and $\sum\limits_{\alpha }y_{\alpha }^{2}=1,$ where $A^{\alpha }=M^{1/2}\Phi
^{T}s^{\alpha }\Phi M^{1/2}$. The infimum of equation (15) is given by $%
\underset{z\in \mathbf{C}}{max}\lambda _{1}(z)-\underset{i>1}{\sum }\lambda
_{i}(z)$ analogous to Ref.[8], with $\lambda _{i}(z)$s are the singular
values, in decreasing order, of the matrix $\sum\limits_{\alpha }z_{\alpha
}A^{\alpha }$. $C(\rho )$ is as well expressed by
\begin{equation}
C(\rho )=\max \{0,\underset{z\in \mathbf{C}}{max}\lambda _{1}(z)-\underset{%
i>1}{\sum }\lambda _{i}(z)\}.
\end{equation}%
One can easily see that $C(\rho )=0$ provides a necessary and even
sufficient condition of full separability for tripartite mixed qubit
systems, hence an effective separability criterion. However, it is so
unfortunate that $C(\rho )$ can not serve as a good entanglement measure,
but only an effective criterion to detect whether a state is fully
separable, because $\boldsymbol{C}(\psi )$ for pure states is not invariant
under local unitary transformations.

\section{Examples}

At first, consider the complementary states to SHIFTS\ UPB [19]. SHIFTS\ UPB
is the set of the following four product states%
\begin{equation}
\{\left\vert 0,1,+\right\rangle ,\left\vert 1,+,0\right\rangle ,\left\vert
+,0,1\right\rangle ,\left\vert -,-,-\right\rangle \}
\end{equation}%
with $\pm =(\left\vert 0\right\rangle \pm \left\vert 1\right\rangle )/\sqrt{2%
}$. The corresponding bound entangled (complementary) state is given by
\begin{equation}
\bar{\rho}=\frac{1}{4}\left( 1-\sum_{i=1}^{4}\left\vert \psi
_{i}\right\rangle \left\langle \psi _{i}\right\vert \right)
\end{equation}%
with $\{\psi _{i}:i=1,\cdot \cdot \cdot ,4\}$ corresponding to the
SHIFTS\ UPB. In Ref.[19], it is stated that this complementary
state has the curious property that not only is it two-way ppt,
it\ is also two-way separable. The numerical
result based on our criterion can show a \textit{non-zero} ($0.1469$) entanglement for $%
C(\bar{\rho})$, which is consistent to [19].

Let's consider the second example, the D\"{u}r-Cirac -Tarrach states [20]
\begin{equation}
\rho _{DCT}=\left(
\begin{array}{cccccccc}
\frac{a+b}{2} & 0 & 0 & 0 & 0 & 0 & 0 & \frac{a-b}{2} \\
0 & c & 0 & 0 & 0 & 0 & 0 & 0 \\
0 & 0 & d & 0 & 0 & 0 & 0 & 0 \\
0 & 0 & 0 & e & 0 & 0 & 0 & 0 \\
0 & 0 & 0 & 0 & e & 0 & 0 & 0 \\
0 & 0 & 0 & 0 & 0 & d & 0 & 0 \\
0 & 0 & 0 & 0 & 0 & 0 & c & 0 \\
\frac{a-b}{2} & 0 & 0 & 0 & 0 & 0 & 0 & \frac{a+b}{2}%
\end{array}%
\right) ,
\end{equation}%
we can also show the $nozero$ $C(\rho _{DCT})$ ($0.3747$) for $a=\frac{1}{3}%
;c=d=\frac{1}{6};b=e=0.$The conclusion is also implied in [20].

The above numerical tests are operated as follows. In order to show the
nonzero $C(x)$ with $x=\bar{\rho}$ or $\rho _{DCT}$, we choose $10^{5}$
random vectors $\left( z_{1},z_{2},\cdot \cdot \cdot ,z_{9}\right) $
generated by \textit{Matlab 6.5 }for a given $x$, then substitute these
vectors into $\sum\limits_{\alpha }z_{\alpha }A^{\alpha }$ and obtain $%
10^{5} $ matrices. We can get $10^{5}$ ($\lambda _{1}(z)-\underset{i>1}{\sum
}\lambda _{i}(z)$)s by singular value decomposition for the matrices. The
maximal $\lambda _{1}(z)-\underset{i>1}{\sum }\lambda _{i}(z)$ among the
matrices is assigned to $C(x)$. Due to the whole process, it is obvious that
our numerical approach is more effective to test the \textit{nonzero} $%
C(\rho )$. It can only provide a reference for the \textit{zero} $C(x)$. If
a standard numerical process is needed, we suggest that the approach
introduced in Ref.[17] be preferred.

\section{\protect\bigskip Conclusion and discussion}

As a summary, we have shown effective criteria for tripartite qubit systems
by the pioneering application of the approach to define a tripartite pure
state as a three-order tensor. However, although our criteria can be reduced
to Wootters' concurrence [7] for bipartite systems, as mentioned above, the
criteria can not serve as a good entanglement measure. Therefore, it is not
necessary to find out the concrete value of $C(\rho )$, but whether $C(\rho
) $ are greater than $zero$, as can be found in our examples. Based on the
tensor treatment for a tripartite pure state, if a more suitable $%
\boldsymbol{C}(\psi )$ that can serve as a good entanglement measure can be
found, it will be interesting. It deserves our attention that our approach
can be easily extended to test the full separability of multipartite systems
in arbitrary dimension, which will be given out in the forthcoming works.

\section{Acknowlegdement}

We would like to thank X. X. Yi for extensive and valuable advice. We are
grateful to the referees for their useful suggestion and comments. This work
was supported by Ministry of Science and Technology, China, under grant
No.2100CCA00700.


\begin{thebibliography}{99}
\bibitem{[21]} Different from the previous definition of tensors, all the
quantities with indices, such as $T_{ijk}$, are called three-order tensors
here. Therefore, the set of all one-order tensor is the set of vectors, and
the set of all two-order tensors is the set of matrices. Three-order tensors
$T_{ijk}$ are matrices correpsonding to the planes in the cube (vectors
corresponding to the edges) when any one (two) of their three indices is
(are) fixed.
\end{thebibliography}
\end{document}